\documentclass{ws-ijmpd}

\begin{document}

\markboth{J. W. Moffat}{A modified gravity and its consequences
for the solar system, astrophysics and cosmology}

%
\catchline{}{}{}{}{}
%

\title{A MODIFIED GRAVITY AND ITS CONSEQUENCES FOR
THE SOLAR SYSTEM, ASTROPHYSICS AND COSMOLOGY\footnote{Talk
presented at the International Workshop {\it From Quantum to
Cosmos: Fundamental Physics in Space}, 22-24 May, 2006, Warrenton,
Virginia, USA}}

\author{J. W. MOFFAT}

\address{Perimeter Institute for Theoretical Physics, \\
31 Caroline St. North, Waterloo, Ontario, N2L 2Y5, Canada}

\address{Department of Physics, University of Waterloo, Address\\Waterloo,
Ontario, N2L 3G1, Canada \\
john.moffat@utoronto.ca}

\maketitle

\begin{history}
\received{9 August 2006} \revised{Day Month Year} \comby{Managing
Editor}
\end{history}

\begin{abstract}
A relativistic modified gravity (MOG) theory leads to a
self-consistent, stable gravity theory that can describe the solar
system, galaxy and clusters of galaxies data and cosmology.
\end{abstract}


\section{Introduction}

A relativistic modified gravity (MOG) called Scalar-Tensor-Vector
Gravity (STVG) desribes a self-consistent, stable gravity theory
that contains Einstein's general relativity in a well-defined
limit.\cite{Moffat} The theory has an extra degree of freedom, a
vector field called a ``phion'' field whose curl is a skew
symmetric field that couples to matter (``fifth force''). The
spacetime geometry is described by a symmetric Einstein metric. An
alternative relativistic gravity theory called Metric-Skew-Tensor
Gravity (MSTG) has also been formulated\cite{Moffat2} in which
the spacetime is described by a symmetric metric, and the extra
degree of freedom is a skew symmetric second rank tensor field.
Both of these theories yield the same weak field consequences for
physical systems.

The classical STVG theory allows the gravitational coupling
``constant'' $G$ and the coupling of the phion field and its
effective mass to vary with space and time as scalar fields.

A MOG should explain the following physical phenomena:

\begin{enumerate}

\item Galaxy rotation curve data;

\item Mass profiles of x-ray clusters;

\item Gravitational lensing data for galaxies and clusters of
galaxies;

\item The cosmic microwave background (CMB) including the
acoustical oscillation power spectrum data;

\item The formation of proto-galaxies in the early universe and
the growth of galaxies;

\item N-body simulations of galaxy surveys;

\item The accelerating expansion of the universe.

\end{enumerate}

We seek a unified description of solar system, astrophysical and
large-scale cosmological data without exotic non-baryonic dark
matter. Dark matter in the form of particles has until now not
been discovered in spite of large-scale experimental
efforts.\cite{Baudis} The accelerating expansion of the universe
should be explained by the MOG theory without postulating a
cosmological constant.

\section{Action and Field Equations}

Our MOG action takes the form\cite{Moffat}:
\begin{equation}
S=S_{\rm Grav}+S_\phi+S_S+S_M,
\end{equation}
where
\begin{equation}
S_{\rm Grav}=\frac{1}{16\pi}\int d^4x\sqrt{-g}\biggl[
\frac{1}{G}(R+2\Lambda)\biggr],
\end{equation}
\begin{equation}
S_\phi=-\int
d^4x\sqrt{-g}\biggl[\omega\biggl(\frac{1}{4}B^{\mu\nu}B_{\mu\nu} +
V(\phi)\biggr)\biggr],
\end{equation}
and
\begin{equation}
\label{Saction} S_S=\int d^4x\sqrt{-g}({\cal F}_1+{\cal F}_2+{\cal
F}_3),
\end{equation}
where
\begin{equation}
{\cal F}_1=\frac{1}{G^3}\biggl(\frac{1}{2}g^{\mu\nu}\nabla_\mu
G\nabla_\nu G-V(G)\biggr),
\end{equation}
\begin{equation}
{\cal F}_2=\frac{1}{G}\biggl(\frac{1}{2}g^{\mu\nu}\nabla_\mu\omega
\nabla_\nu\omega-V(\omega)\biggr),
\end{equation}
\begin{equation}
{\cal F}_3
=\frac{1}{\mu^2G}\biggl(\frac{1}{2}g^{\mu\nu}\nabla_\mu\mu\nabla_\nu\mu
-V(\mu)\biggr).
\end{equation}

We have chosen units with $c=1$, $\nabla_\mu$ denotes the
covariant derivative with respect to the metric $g_{\mu\nu}$. We
adopt the metric signature $\eta_{\mu\nu}={\rm diag}(1,-1,-1,-1)$
where $\eta_{\mu\nu}$ is the Minkowski spacetime metric, and
$R=g^{\mu\nu}R_{\mu\nu}$ where $R_{\mu\nu}$ is the symmetric Ricci
tensor. Moreover, $V(\phi)$ denotes a potential for the vector
field $\phi^\mu$, while $V(G), V(\omega)$ and $V(\mu)$ denote the
three potentials associated with the three scalar fields
$G(x),\omega(x)$ and $\mu(x)$, respectively. The field $\omega(x)$
is dimensionless and $\Lambda$ denotes the cosmological constant.
Moreover,
\begin{equation}
B_{\mu\nu}=\partial_\mu\phi_\nu-\partial_\nu\phi_\mu.
\end{equation}
The field equations and the test particle equations of motion are
derived in Ref.~\refcite{Moffat}.

The action for the field $B_{\mu\nu}$ is of the Maxwell-Proca form
for a massive vector field $\phi_\mu$. It can be proved that this
MOG possesses a stable vacuum and the Hamiltonian is bounded from
below. Even though the action is not gauge invariant, it can be
shown that the longitudinal mode $\phi_0$ (where
$\phi_\mu=(\phi_0,\phi_i)\, (i=1,2,3)$) does not propagate and the
theory is free of ghosts. Similar arguments apply to the MSTG
theory.\cite{Moffat2}\footnote{For a detailed discussion of
possible instabilities and pathological behavior of vector-gravity
theories, see Ref.~\refcite{Clayton}.}

\section{Modified Newtonian Acceleration Law and Galaxy Dynamics}

The modified acceleration law can be written as\cite{Moffat}:
\begin{equation}
\label{accelerationGrun} a(r)=-\frac{G(r)M}{r^2},
\end{equation}
where
\begin{equation}
\label{runningG}
G(r)=G_N\biggl[1+\sqrt{\frac{M_0}{M}}\biggl(1-\exp(-r/r_0)
\biggl(1+\frac{r}{r_0}\biggr)\biggr)\biggr]
\end{equation}
is an {\it effective} expression for the variation of $G$ with
respect to $r$, and $G_N$ denotes Newton's gravitational constant.
A good fit to a large number of galaxies has been achieved with
the parameters\cite{Brownstein}:
\begin{equation}
M_0=9.60\times 10^{11}\,M_{\odot},\quad r_0=13.92\,{\rm
kpc}=4.30\times 10^{22}\,{\rm cm}.
\end{equation}
In the fitting of the galaxy rotation curves for both LSB and HSB
galaxies, using photometric data to determine the mass
distribution ${\cal M}(r)$,\cite{Brownstein} only the
mass-to-light ratio $\langle M/L\rangle$ is employed, once the
values of $M_0$ and $r_0$ are fixed universally for all LSB and
HSB galaxies. Dwarf galaxies are also fitted with the
parameters\cite{Brownstein}:
\begin{equation}
M_0=2.40\times 10^{11}\,M_{\odot},\quad r_0=6.96\,{\rm
kpc}=2.15\times 10^{22}\,{\rm cm}.
\end{equation}
By choosing universal values for the parameters
$G_{\infty}=G_N(1+\sqrt{M_0/M})$, $(M_0)_{\rm clust}$ and
$(r_0)_{\rm clust}$, we are able to obtain satisfactory fits to a
large sample of X-ray cluster data.\cite{Brownstein2}

\section{Solar System and Binary Pulsar}

Let us assume that we are in a distance scale regime for which the
fields $G$, $\omega$ and $\mu$ take their approximate renormalized
constant values:
\begin{equation}
\label{constants} G\sim G_0(1+Z),\quad \omega\sim\omega_0A,\quad
\mu\sim\mu_0B,
\end{equation}
where $G_0, \omega_0$ and $\mu_0$ denote the ``bare'' values of
$G, \omega$ and $\mu$, respectively, and $Z, A$ and $B$ are the
associated renormalization constants. We obtain from the equations
of motion of a test particle the orbital equation (we reinsert the
speed of light c)\cite{Moffat}:
\begin{equation}
\label{particleorbit} \frac{d^2u}{d\phi^2}+u=\frac{GM}{c^2
J^2}-\frac{K}{c^2J^2}\exp(-r/r_0)\biggl[1
+\biggl(\frac{r}{r_0}\biggr)\biggr]+\frac{3GM}{c^2}u^2.
\end{equation}
where $u=1/r$, $K=G_N\sqrt{MM_0}$ and $J$ denotes the orbital
angular momentum. Using the large $r$ weak field approximation, we
obtain the orbit equation for $r\ll r_0$:
\begin{equation}
\label{orbitperihelion}
\frac{d^2u}{d\phi^2}+u=N+3\frac{GM}{c^2}u^2,
\end{equation}
where $J_N$ denotes the Newtonian value of $J$ and
\begin{equation}
N=\frac{GM}{c^2J_N^2}-\frac{K}{c^2J_N^2}.
\end{equation}

We can solve Eq.(\ref{orbitperihelion}) by perturbation theory and
find for the perihelion advance of a planetary orbit
\begin{equation}
\label{perihelionformula} \Delta\omega=\frac{6\pi}{c^2L}
(GM_{\odot}-K_{\odot}),
\end{equation}
where $J_N=(GM_{\odot}L/c^2)^{1/2}$, $L=a(1-e^2)$ and $a$ and $e$
denote the semimajor axis and the eccentricity of the planetary
orbit, respectively.

For the solar system $r\ll r_0$ and from the running of the
effective gravitational coupling constant, $G=G(r)$, we have
$G\sim G_N$ within the experimental errors for the measurement of
Newton's constant $G_N$. We choose for the solar system
\begin{equation}
\label{perhbound} \frac{K_{\odot}}{c^2}\ll 1.5\,{\rm km}
\end{equation}
and use $G=G_N$ to obtain from (\ref{perihelionformula}) a
perihelion advance of Mercury in agreement with GR. The bound
(\ref{perhbound}) requires that the coupling constant $\omega$
varies with distance in such a way that it is sufficiently small
in the solar system regime and determines a value for $M_0$ that
is in accord with the bound (\ref{perhbound}).

For terrestrial experiments and orbits of satellites, we have also
that $G\sim G_N$ and for $K_{\oplus}$ sufficiently small, we then
achieve agreement with all gravitational terrestrial experiments
including E\"otv\"os free-fall experiments and ``fifth force''
experiments.

For the binary pulsar PSR 1913+16 the formula
(\ref{perihelionformula}) can be adapted to the periastron shift
of a binary system. Combining this with the STVG gravitational
wave radiation formula, which will approximate closely the GR
formula, we can obtain agreement with the observations for the
binary pulsar.  The mean orbital radius for the binary pulsar is
equal to the projected semi-major axis of the binary, $\langle
r\rangle_N=7\times 10^{10}\,{\rm cm}$, and we choose $\langle
r\rangle_N\ll r_0$. Thus, for $G=G_N$ within the experimental
errors, we obtain agreement with the binary pulsar data for the
periastron shift when
\begin{equation}
\label{binarybound} \frac{K_N}{c^2}\ll 4.2\,{\rm km}.
\end{equation}

For a massless photon we have
\begin{equation}
\label{lightbending} \frac{d^2u}{d\phi^2}+u=3\frac{GM}{c^2}u^2.
\end{equation}
For the solar system using $G\sim G_N$ within the experimental
errors gives the light deflection:
\begin{equation}
\Delta_{\odot}=\frac{4G_NM_{\odot}}{c^2R_{\odot}}
\end{equation}
in agreement with GR.

\section{Pioneer Anomaly}

The radio tracking data from the Pioneer 10/11 spacecraft during
their travel to the outer parts of the solar system have revealed
an anomalous acceleration. The Doppler data obtained at distances
$r$ from the Sun between $20$ and $70$ astronomical units (AU)
showed the anomaly as a deviation from Newton's and Einstein's
gravitational theories. The anomaly is observed in the Doppler
residuals data, as the differences of the observed Doppler
velocity from the modelled Doppler velocity, and can be
represented as an anomalous acceleration directed towards the Sun,
with an approximately constant amplitude over the range of
distance, $20\, {\rm au} < r < 70\, {\rm au}$\cite{Anderson,Anderson2,Turyshev,Anderson3}:
\begin{equation} \label{aP}
a_P=(8.74\pm 1.33)\times 10^{-8}\,{\rm cm}\,s^{-2}.
\end{equation}
After a determined attempt to account for all {\it known} sources
of systematic errors, the conclusion has been reached that the
anomalous acceleration towards the Sun could be a real physical
effect that requires a physical
explanation.\cite{Anderson,Anderson2,Turyshev,Anderson3}\footnote{It
is possible that a heat transfer mechanism from the spacecraft
transponders could produce a non-gravitational explanation for the
anomaly.}

We can rewrite the acceleration in the form
\begin{equation}
\label{accelerationlaw} a(r)=-\frac{G_N
M}{r^2}\biggl\{1+\alpha(r)\biggl[1-\exp(-r/\lambda(r))
\biggl(1+\frac{r}{\lambda(r)}\biggr)\biggr]\biggr\}.
\end{equation}

We postulate a gravitational solution that the Pioneer 10/11
anomaly is caused by the difference between the running of $G(r)$
and the Newtonian value, $G_N$. So the Pioneer anomalous
acceleration directed towards the center of the Sun is given by
\begin{equation}
a_P=-\frac{\delta G(r)M_{\odot}}{r^2},
\end{equation}
where
\begin{equation} \label{deltaG}
\delta G(r)=G_N\alpha(r)\biggl[1-\exp(-r/\lambda(r))
\biggl(1+\frac{r}{\lambda(r)}\biggr)\biggr].
\end{equation}
Lacking at present a solution for the variations of $\alpha(r)$
and $\lambda(r)$ in the solar system, we adopt the following
parametric representations of the ``running'' of $\alpha(r)$ and
$\lambda(r)$:
\begin{equation}
\label{alpha} \alpha(r) =\alpha_\infty(1-\exp(-r/{\bar r}))^{b/2},
\end{equation}
\begin{equation}
\label{lambda} \lambda(r)=\lambda_\infty{(1-\exp(-r/{\bar
r}))^{-b}}.
\end{equation}
Here, ${\bar r}$ is a non-running distance scale parameter and $b$
is a constant.

In Ref.~\refcite{Brownstein3}, a best fit to the acceleration data
extracted from Figure 4 of~Ref.~\refcite{Anderson3} was obtained using a
nonlinear least-squares fitting routine including estimated errors
from the Doppler shift observations\cite{Anderson2}. The best fit
parameters are
\begin{eqnarray}
\nonumber \alpha_\infty &=& (1.00\pm0.02)\times 10^{-3},\\
\nonumber \lambda_\infty &=& 47\pm 1\, {\rm au} ,\\
\nonumber {\bar r} &=& 4.6\pm 0.2\, {\rm au} ,\\
\label{bestparameters} b &=& 4.0.
\end{eqnarray}
The small uncertainties in the best fit parameters are due to the
remarkably low variance of residuals corresponding to a reduced
$\chi^{2}$ per degree of freedom of 0.42 signalling a good fit. An
important result obtained from our fit to the anomalous
acceleration data is that the anomalous acceleration kicks-in at
the orbit of Saturn.

Fifth force experimental bounds plotted for $\log_{10}\alpha$
versus $\log_{10}\lambda$ are shown in Fig. 1 of Ref.~\refcite{Reynaud}
for fixed values of $\alpha$ and $\lambda$. The updated 2003
observational data for the bounds obtained from the planetary
ephemerides is extrapolated to $r = 10^{15}\,\mbox{m}=6,685\, {\rm
au}$\cite{Fischbach2}. However, this extrapolation is based on
using fixed universal values for the parameters $\alpha$ and
$\lambda$. Since known reliable data from the ephemerides of the
outer planets ends with the data for Pluto at a distance from the
Sun, $r=39.52\, {\rm au}=5.91\times 10^{12}\,m$, we could claim
that for our range of values $47\, {\rm au} < \lambda(r) <
\infty$, we predict $\alpha(r)$ and $\lambda(r)$ values consistent
with the {\it un-extrapolated} fifth force bounds.

A consequence of a variation of $G$ and $GM_\odot$ for the solar
system is a modification of Kepler's third law:
\begin{equation}
\label{Kepler}
a_{PL}^3=G(a_{PL})M_\odot\biggl(\frac{T_{PL}}{2\pi}\biggr)^2,
\end{equation}
where $T_{PL}$ is the planetary sidereal orbital period and
$a_{PL}$ is the physically measured semi-major axis of the
planetary orbit. For given values of $a_{PL}$ and $T_{PL}$,
(\ref{Kepler}) can be used to determine $G(r)M_\odot$.

For several planets such as Mercury, Venus, Mars and Jupiter there
are planetary ranging data, spacecraft tracking data and
radiotechnical flyby observations available, and it is possible to
measure $a_{PL}$ directly. For a distance varying $GM_\odot$ we
derive\cite{Fischbach,Talmadge}:
\begin{equation}
\label{eta0} \biggl(\frac{a_{PL}}{{\bar
a}_{PL}}\biggr)=1+\eta_{PL}
=\biggl[\frac{G(a_{PL})M_\odot}{G(a_\oplus)M_\odot}\biggr]^{1/3}.
\end{equation}
Here, it is assumed that $GM_\odot$ varies with distance such that
$\eta_{PL}$ can be treated as a constant for the orbit of a
planet.  We obtain
\begin{equation}
\label{eta} \eta_{PL}
=\left[\frac{G(a_{PL})}{G(a_{\oplus})}\right]^{1/3}-1.
\end{equation}

The results for $\Delta \eta_{PL}$ due to the uncertainty in the
planetary ephemerides are presented in Ref.~\refcite{Brownstein3} for
the nine planets and are consistent with the solar ephemerides.

The validity of the bounds on a possible fifth force obtained from
the ephemerides of the outer planets Uranus, Neptune and Pluto are
critical in the exclusion of a parameter space for our fits to the
Pioneer anomaly acceleration. Beyond the outer planets, the
theoretical prediction for $\eta(r)$ approaches an asymptotic
value:
\begin{equation} \label{etalim}
\eta_{\infty} \equiv \lim_{r \to \infty} \eta(r)= 3.34 \times
10^{-4}.
\end{equation}
We see that the variations (``running'') of $\alpha(r)$ and
$\lambda(r)$ with distance play an important role in interpreting
the data for the fifth force bounds. This is in contrast to the
standard non-modified Yukawa correction to the Newtonian force law
with fixed universal values of $\alpha$ and $\lambda$ and for the
range of values $0 < \lambda < \infty$, for which the equivalence
principle and lunar laser ranging and radar ranging data to
planetary probes exclude the possibility of a gravitational and
fifth force explanation for the Pioneer
anomaly.\cite{Adelberger,Adelberger2,Will}

A study of the Shapiro time delay prediction in our MOG is found
to be consistent with time delay observations and predicts a
measurable deviation from GR for the outer planets Neptune and
Pluto.\cite{Moffat3}

\section{Gravitational Lensing}

The bending angle of a light ray as it passes near a massive
system along an approximately straight path is given to lowest
order in $v^2/c^2$ by
\begin{equation}
\label{lensingformula} \theta=\frac{2}{c^2}\int\vert
a^{\perp}\vert dz,
\end{equation}
where $\perp$ denotes the perpendicular component to the ray's
direction, and dz is the element of length along the ray and $a$
denotes the acceleration.

From (\ref{lightbending}), we obtain the light deflection
\begin{equation}
\Delta=\frac{4GM}{c^2R}=\frac{4G_N{\overline M}}{c^2R},
\end{equation}
where
\begin{equation}
{\overline M}=M\biggl(1+\sqrt{\frac{M_0}{M}}\biggr).
\end{equation}
The value of ${\overline M}$ follows from (\ref{runningG}) for
clusters as $r\gg r_0$ and
\begin{equation}
G(r)\rightarrow
G_{\infty}=G_N\biggl(1+\sqrt{\frac{M_0}{M}}\biggr).
\end{equation}
We choose for a cluster $M_0=3.6\times 10^{15}\,M_{\odot}$ and a
cluster mass $M_{\rm clust}\sim 10^{14}\,M_{\odot}$, and obtain
\begin{equation}
\biggl(\sqrt{\frac{M_0}{M}}\biggr)_{\rm clust}\sim 6.
\end{equation}
We see that ${\overline M}\sim 7M$ and we can explain the increase
in the light bending without exotic dark matter.

For $r\gg r_0$ we get
\begin{equation}
a(r)=-\frac{G_N\overline M}{r^2}.
\end{equation}
We expect to obtain from this result a satisfactory description of
lensing phenomena using Eq.(\ref{lensingformula}).

\section{Modified Friedmann Equations in Cosmology}

We shall base our results for the cosmic microwave background
(CMB) power spectrum on our MOG without a second component of cold
dark matter (CDM). Our description of the accelerating
universe\cite{Perlmutter,Riess} is based on ${\Lambda}_G$ in
Eq.(\ref{LambdaG}) derived from our varying gravitational
constant.\cite{Moffat4}

We adopt a homogeneous and isotropic
Friedmann-Lema\^{i}tre-Robertson-Walker (FLRW) background geometry
with the line element
\begin{equation}
ds^2=dt^2-a^2(t)\biggl(\frac{dr^2}{1-kr^2}+r^2d\Omega^2\biggr),
\end{equation}
where $d\Omega^2=d\theta^2+\sin^2\theta d\phi^2$ and $k=0,-1,+1$
for a spatially flat, open and closed universe, respectively. Due
to the symmetry of the FLRW background spacetime, we have
$\phi_0\equiv\phi\not= 0$, $\phi_i=0$ and $B_{\mu\nu}=0$.

We define the energy-momentum tensor for a perfect fluid by
\begin{equation}
T^{\mu\nu}=(\rho+p)u^\mu u^\nu-pg^ {\mu\nu},
\end{equation}
where $u^\mu=dx^\mu/ds$ is the 4-velocity of a fluid element and
$g_{\mu\nu}u^\mu u^\nu=1$. Moreover, we have
\begin{equation}
\rho=\rho_m+\rho_\phi+\rho_S,\quad p=p_m+p_\phi+p_S,
\end{equation}
where $\rho_i$ and $p_i$ denote the components of density and
pressure associated with the matter, the $\phi^\mu$ field and the
scalar fields $G$, $\omega$ and $\mu$, respectively.

The modified Friedmann equations take the form\cite{Moffat}:
\begin{equation}
\label{Friedmann1} \frac{\dot
a^2(t)}{a^2(t)}+\frac{k}{a^2(t)}=\frac{8\pi
G(t)\rho(t)}{3}+f(t)+\frac{\Lambda}{3},
\end{equation}
\begin{equation}
\label{Friedmann2} \frac{{\ddot a}(t)}{a(t)}=-\frac{4\pi
G(t)}{3}[\rho(t)+3p(t)]+h(t) +\frac{\Lambda}{3},
\end{equation}
where $\dot a=da/dt$ and
\begin{equation}
\label{fequation} f(t)=\frac{\dot a(t)}{a(t)}\frac{{\dot
G(t)}}{G(t)},
\end{equation}
\begin{equation}
\label{hequation} h(t)=\frac{1}{2}\biggl(\frac{{\ddot
G(t)}}{G(t)}-\frac{{{\dot G}^2(t)}}{G^2(t)}+2\frac{\dot
a(t)}{a(t)}\frac{{\dot G(t)}}{G(t)}\biggr).
\end{equation}
From (\ref{Friedmann1}) we obtain
\begin{equation}
\rho a^3=\frac{3}{8\pi G}a\biggl({\dot
a}^2+k-a^2f-\frac{1}{3}a^2\Lambda\biggr).
\end{equation}
This leads by differentiation with respect to $t$ to the
expression:
\begin{equation}
\label{conserveeq} \dot\rho+3\frac{d\ln a}{dt}(\rho+p)+{\cal I}=0,
\end{equation}
where
\begin{equation}
\label{Iequation} {\cal I}=\frac{3a^2}{8\pi G}(2{\dot a}f+a\dot
f-2{\dot a}h).
\end{equation}

An approximate solution to the field equations for the variation
of $G$ in Ref.~\refcite{Moffat} in the background FLRW spacetime is
given by
\begin{equation}
\label{approxGeq} {\ddot{\cal G}}+3H{\dot{\cal G}}+V'({\cal
G})=\frac{1}{2}G_N{\cal G}^2\biggl(\rho-3p+\frac{\Lambda}{4\pi
G_N{\cal G}}\biggr),
\end{equation}
where ${\cal G}(t)=G(t)/G_N$ and $H=\dot a/a$. A solution for
${\cal G}$ in terms of a given potential $V({\cal G})$ and for
given values of $\rho$, $p$ and $\Lambda$ can be obtained from
(\ref{approxGeq}).\cite{Moffat4}

The solution for ${\cal G}$ must satisfy a constraint at the time
of big bang nucleosynthesis.\cite{Bean} The number of
relativistic degrees of freedom is very sensitive to the cosmic
expansion rate at 1 MeV. This can be used to constrain the time
dependence of $G$. Measurements of the $^4He$ mass fraction and
the deuterium abundance at 1 MeV lead to the constraint $G(t)\sim
G_N$. We impose the condition ${\cal G}(t)\rightarrow 1$ as
$t\rightarrow t_{BBN}$ where $t_{BBN}$ denotes the time of the big
bang nucleosynthesis. Moreover, {\it locally in the solar system}
we must satisfy the observational bound from the Cassini
spacecraft measurements\cite{Bertotti}:
\begin{equation}
\vert{\dot G}/G\vert\leq 10^{-12}{\rm yr}^{-1}.
\end{equation}

We shall now impose the approximate conditions at the epoch of
recombination:
\begin{equation}
\label{fcond} 2{\dot a}f+a\dot f\sim 2{\dot a}h,
\end{equation}
\begin{equation}
\label{Gcond} \frac{d}{dt}\biggl(\frac{\dot G}{G}\biggr) <
2\frac{\dot a}{a}\frac{\dot G}{G}.
\end{equation}
We find from (\ref{hequation}) and (\ref{Gcond}) that $f\sim h$,
and from the condition (\ref{fcond}) we obtain
\begin{equation}
\dot f\equiv\frac{d\Lambda_G}{dt}\sim 0,
\end{equation}
where
\begin{equation}
\Lambda_G=\frac{\dot a}{a}\frac{{\dot G}}{G}.
\end{equation}

By setting the cosmological constant $\Lambda=0$, we get the
generalized Friedmann equations
\begin{equation}
\label{Fried1} \frac{\dot a^2}{a^2}+\frac{k}{a^2}=\frac{8\pi
G\rho}{3}+\Lambda_G,
\end{equation}
\begin{equation}
\label{Fried2} \frac{{\ddot a}}{a}=-\frac{4\pi
G}{3}(\rho+3p)+\Lambda_G.
\end{equation}

We now have from (\ref{conserveeq}), (\ref{Iequation}) and
(\ref{fcond}) at the epoch of recombination ${\cal I}\sim 0$ and
\begin{equation}
\label{GRconserveeq} \dot\rho+3\frac{d\ln a}{dt}(\rho+p)\sim 0.
\end{equation}
We adopt the equation of state: $p(t)=w\rho(t)$ and derive from
(\ref{GRconserveeq}) the approximate solution for $\rho(t)$:
\begin{equation}
\rho(t)\sim \rho(t_0)\biggl(\frac{a_0}{a(t)}\biggr)^{3(1+w)},
\end{equation}
where $a/a_0=1/(1+z)$ and z denotes the red shift. For the matter
and radiation densities $\rho_m$ and $\rho_r$, we have $w=0$ and
$w=1/3$, respectively. This gives
\begin{equation}
\rho_m(t)\sim\rho_m(t_0)(1+z)^3,\quad \rho_r(t)\sim
\rho_r(t_0)(1+z)^4.
\end{equation}

Let us expand $G(t)$ in a power series
\begin{equation}
G(t)=G_{\rm eff}(t_r)+(t-t_r){\dot G}(t_r)+(t-t_r)^2{\ddot
G}(t_r)+... .
\end{equation}
where $t\sim t_r$ is the time of recombination and $G_{\rm
eff}(t_r)=G_N(1+Z)={\rm const.}$ We write the generalized
Friedmann equation for flat space, $k=0$, in the approximate form
\begin{equation}
\label{effectFried} H^2=\frac{8\pi G_{\rm
eff}\rho_m}{3}+\Lambda_G,
\end{equation}
where
\begin{equation}
\label{LambdaG} \Lambda_G=H\frac{{\dot G}}{G} > 0
\end{equation}
and ${\dot\Lambda}_G\sim 0$. It follows from (\ref{effectFried})
that for a spatially flat universe:
\begin{equation}
\label{omegaeq} \Omega_m+\Omega_{G}=1,
\end{equation}
where
\begin{equation}
\label{Omegaeq} \Omega_m=\frac{8\pi G_{\rm eff}\rho_m}{3
H^2},\quad \Omega_G=\frac{\Lambda_G}{H^2}.
\end{equation}
We shall postulate that the matter density $\rho_m$ is dominated
by the baryon density, $\rho_m\sim\rho_b$, and we have
\begin{equation}
\Omega_m\sim\Omega_{b{\rm eff}},
\end{equation}
where
\begin{equation}
\Omega_{b{\rm eff}}=\frac{8\pi G_{\rm eff}\rho_b}{3 H^2}.
\end{equation}
Thus, we assume that {\it the baryon-photon fluid dominates matter
before recombination and at the surface of last scattering without
a cold dark matter fluid component}.

From the current value: $H_0=7.5\times 10^{-11}\,{\rm yr}^{-1}$
and (\ref{LambdaG}) and (\ref{Omegaeq}), we obtain for
$\Omega_G\sim 0.7$:
\begin{equation}
\label{scaledG} \vert {\dot G}/{G}\vert\sim 5\times 10^{-11}\,{\rm
yr}^{-1},
\end{equation}
valid at cosmological scales for red shifts $ z > 0.1$. In the
{\it local solar system} and for the binary pulsar PSR 1913+16 for
$z\sim 0$, the experimental bound is
\begin{equation}
\vert {\dot G}/{G}\vert < 5\times 10^{-12}\,{\rm yr}^{-1}.
\end{equation}
We can explain the accelerated expansion of the universe deduced
from supernovae measurements in the range $0.1 < z < 1.7$ using
the cosmologically scaled value of ${\dot G}/{G}$ in
(\ref{scaledG}) with Einstein's cosmological constant $\Lambda=0$.

\section{Acoustical Peaks in the CMB Power Spectrum}

Mukhanov\cite{Mukhanov} has obtained an analytical solution to
the amplitude of fluctuations in the CMB power spectrum for $l\gg
1$:
\begin{equation}
\label{fluctuations} l(l+1)C_l\sim \frac{B}{\pi}(O+N).
\end{equation}
Here, $O$ denotes the oscillating part of the spectrum, while the
non-oscillating contribution can be written as the sum of three
parts
\begin{equation}
N=N_1+N_2+N_3.
\end{equation}

The oscillating contributions can be calculated from the formula
\begin{equation}
O\sim
\sqrt{\frac{\pi}{r_hl}}\biggl[A_1\cos\biggl(lr_p+\frac{\pi}{4}\biggr)
+A_2\cos\biggl(2lr_p+\frac{\pi}{4}\biggr)\biggr]\exp(-(l/l_s)^2),
\end{equation}
where $r_h$ and $r_p$ are parameters that determine predominantly
the heights and positions of the peaks, respectively. The $A_1$
and $A_2$ are constant coefficients given in the range $100 < l <
1200$ for $\Omega_m\sim\Omega_{b{\rm eff}}$ by
\begin{equation}
\label{A1coefficient} A_1\sim 0.1\xi\frac{(({\cal
P}-0.78)^2-4.3)}{(1+\xi)^{1/4}}\exp\biggl(\frac{1}{2}(l_s^{-2}-l_f^{-2})l^2\biggr),
\end{equation}
\begin{equation}
\label{A2coefficient} A_2\sim 0.14\frac{(0.5+0.36{\cal
P})^2}{(1+\xi)^{1/2}},
\end{equation}
where
\begin{equation}
{\cal P}=\ln\biggl(\frac{lI}{200(\Omega_{b\rm eff})^{1/2}}\biggr),
\end{equation}
and $I$ is given by the ratio
\begin{equation}
\frac{\eta_x}{\eta_0}\sim
\frac{I}{z_x^{1/2}}=3\biggl(\frac{\Omega_G}{\Omega_{b\rm
eff}}\biggr)^{1/6} \biggl(\int_0^y\frac{dx}{(\sinh
x)^{2/3}}\biggr)^{-1}\frac{1}{z_x^{1/2}}.
\end{equation}
Here, $\eta_x$ and $z_x$ denote a conformal time $\eta=\eta_x$ and
a redshift in the range $\eta_0 > \eta_x > \eta_r$ when radiation
can be neglected and $y=\sinh^{-1}(\Omega_G/\Omega_{b\rm
eff})^{1/2}$. To determine $\eta_x/\eta_0$, we use the exact
solution for a flat dust-dominated universe with a constant
$\Lambda_G$:
\begin{equation}
a(t)=a_0\biggl(\sinh\biggl(\frac{3}{2}\biggr)H_0t\biggr)^{2/3},
\end{equation}
where $a_0$ and $H_0$ denote the present values of $a$ and the
Hubble parameter $H$.

The $l_f$ and $l_s$ in (\ref{A1coefficient}) denote the finite
thickness and Silk damping scales, respectively, given by
\begin{equation}
l_f^2=\frac{1}{2\sigma^2}\biggl(\frac{\eta_0}{\eta_r}\biggr)^2,\quad
l_s^2=\frac{1}{2(\sigma^2+1/(k_D\eta_r)^2)}\biggl(\frac{\eta_0}{\eta_r}\biggr)^2,
\end{equation}
where
\begin{equation}
\sigma\sim 1.49\times 10^{-2}\biggl[1+\biggl(1+\frac{z_{\rm
eq}}{z_r}\biggr)^{-1/2}\biggr],\quad
k_D(\eta)=\biggl(\frac{2}{5}\int_0^\eta d\eta
c^2_s\frac{\tau_\gamma}{a}\biggr)^{-1/2},
\end{equation}
and $\tau_\gamma$ is the photon mean-free time.

A numerical fitting formula gives\cite{Mukhanov}:
\begin{equation}
{\cal P}\sim\ln\biggl(\frac{l}{200(\Omega_{b\rm
eff}^{0.59})}\biggr),\quad r_p=\frac{1}{\eta_0}\int d\eta
c_s(\eta).
\end{equation}
Moreover,
\begin{equation}
\xi\equiv
\frac{1}{3c_s^2}-1=\frac{3}{4}\biggl(\frac{\rho_b}{\rho_\gamma}\biggr),
\end{equation}
where $c_s(\eta)$ is the speed of sound:
\begin{equation}
c_s(\eta)=\frac{1}{\sqrt{3}}\biggl[1+\xi\biggl(\frac{a(\eta)}{a(\eta_r)}\biggr)\biggr]^{-1/2}.
\end{equation}
{\it We note that $\xi$ does not depend on the value of $G_{\rm
eff}$}. For the matter-radiation universe:
\begin{equation}
a(\eta)={\bar
a}\biggl[\biggl(\frac{\eta}{\eta_{*}}\biggr)^2+2\biggl(\frac{\eta}{\eta_{*}}\biggr)\biggr],
\end{equation}
where for radiation-matter equality $z=z_{eq}$:
\begin{equation}
\frac{z_{eq}}{z_r}\sim
\biggl(\frac{\eta_r}{\eta_*}\biggr)^2+2\biggl(\frac{\eta_r}{\eta_*}\biggr),
\end{equation}
and $\eta_{eq}=\eta_*(\sqrt{2}-1)$ follows from ${\bar
a}=a(\eta_{eq})$.

For the non-oscillating parts, we have
\begin{equation}
N_1\sim 0.063\xi^2\frac{({\cal
P}-0.22(l/l_f)^{0.3}-2.6)^2}{1+0.65(l/l_f)^{1.4}}\exp(-(l/l_f)^2),
\end{equation}
\begin{equation}
N_2\sim\frac{0.037}{(1+\xi)^{1/2}}\frac{{\cal
P}-0.22(l/l_s)^{0.3}+1.7)^2}{1+0.65(l/l_f)^{1.4}}\exp(-(l/l_s)^2),
\end{equation}
\begin{equation}
N_3\sim\frac{0.033}{(1+\xi)^{3/2}}\frac{{\cal
P}-0.5(l/l_s)^{0.55}+2.2)^2}{1+2(l/l_s)^2}\exp(-(l/l_s)^2).
\end{equation}

Mukhanov's formula\cite{Mukhanov} for the oscillating spectrum is
given by
\begin{equation}
\label{Mukhanov}  C(l)\equiv\frac{l(l+1)C_l}{[l(l+1)C_l]_{{\rm
low}\,l}}=\frac{100}{9}(O+N),
\end{equation}
where we have normalized the power spectrum by using for a flat
spectrum with a constant amplitude $B$:
\begin{equation}
[l(l+1)C_l]_{{\rm low}\,l}=\frac{9B}{100\pi}.
\end{equation}

We adopt the parameters
\begin{equation}
\label{cosmologparam} \Omega_{bN}\sim 0.04,\quad\Omega_{b\rm
eff}\sim 0.3,\quad\Omega_G\sim 0.7,\quad \xi\sim 0.6,
\end{equation}
and
\begin{equation}
\label{cosmologparam2} r_h=0.03,\quad r_p=0.01\quad l_f\sim
1580,\quad l_s\sim 1100,
\end{equation}
where $\Omega_{bN}=8\pi G_N\rho_b/3H^2$.

The fluctuation spectrum determined by Mukhanov's analytical
formula is displayed in Fig. 1 for the choice of cosmological
parameters given in (\ref{cosmologparam}) and
(\ref{cosmologparam2}). \vskip 0.3 in

\begin{figure}[h]
\vskip -24pt
\centerline{\psfig{file=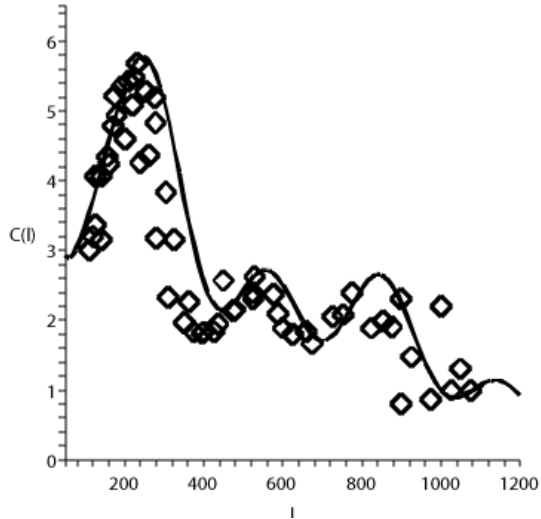,width=7.2cm}}
\caption{The solid line shows the result of the calculation of the
power spectrum acoustical oscillations: $C(l)$, and the $\diamond
s$ correspond to the WMAP, Archeops and Boomerang data in units
$\mu K^2\times 10^{-3}$ as presented in Refs. 27-30.
\label{fig:1}}
\vskip -10pt
\end{figure}

The role played by CDM in the standard scenario is replaced in the
modified gravity theory by the significant deepening of the
gravitational potential well by the effective gravitational
constant, $G_{\rm eff}\sim 7G_N$, that traps the non-relativistic
baryons before recombination. The deepening of the gravitational
well reduces the baryon dissipation due to the photon coupling
pressure and the third and higher peaks in the acoustic
oscillation spectrum are not erased by finite thickness and baryon
drag effects. The effective baryon density $\Omega_{b\rm
eff}=(1+Z)\Omega_{bN}\sim 7\Omega_{bN}\sim 0.3$ dominates the
fluid before recombination, {\it and we fit the acoustical power
spectrum data without a cold dark matter fluid component.} For $t
< t_{\rm dec}$, where $t_{\rm dec}$ denotes the time of
matter-radiation decoupling, luminous baryons and photons are
tightly coupled and for photons the dominant collision mechanism
is scattering by non-relativistic electrons due to Thompson
scattering. It follows that luminous baryons are dragged along
with photons and perturbations at wavelength $\lambda_w < \ell_s$
will be partly erased where $\ell_s$ is the proper Silk length
given by $\ell_s\sim 3.5\,{\rm Mpc}\,\Omega_{b\rm
eff}^{-1/2}$.\cite{Silk} We have $\ell_s\sim 6\,{\rm Mpc}$ for
$\Omega_{b\rm eff}\sim 0.3$ compared to $\ell_s\sim 18\,{\rm Mpc}$
for $\Omega_{bN}\sim 0.04$. The Silk mass is reduced by more than
an order of magnitude\footnote{Note that there will be a fraction
of dark baryonic matter before decoupling.}. Thus, sufficient
baryonic perturbations should survive before $t\sim t_{\rm dec}$
to explain the power spectrum without collisionless dark matter.

Our predictions for the CMB power spectrum for large angular
scales corresponding to $l < 100$ will involve the integrated
Sachs-Wolfe contributions obtained from the modified gravitational
potential.

\section{Conclusions}

We have demonstrated that a modified gravity theory\cite{Moffat}
can lead to a satisfactory fit to the galaxy rotation curve data,
mass profiles of x-ray cluster data, the solar system and the
binary pulsar PSR 1913+16 data. Moreover, we can provide an
explanation for the Pioneer 10/11 anomalous acceleration data,
given that the anomaly is caused by gravity. We can fit
satisfactorily the acoustical oscillation spectrum obtained in the
cosmic microwave background data by employing the analytical
formula for the fluctuation spectrum derived by
Mukhanov.\cite{Mukhanov}

$\Lambda_G$ obtained from the varying gravitational constant in
our MOG replaces the standard cosmological constant $\Lambda$ in
the concordance model. Thus, the accelerating expansion of the
universe is obtained from the MOG scenario.

An important problem to investigate is whether an N-body
simulation calculation based on our MOG scenario can predict the
observed large scale galaxy surveys. The formation of proto-galaxy
structure before and after the epoch of recombination and the
growth of galaxies and clusters of galaxies at later times in the
expansion of the universe has to be explained.

We have succeeded in fitting in a unified picture a large amount
of data over 16 orders of magnitude in distance scale from Earth
to the surface of last scattering some 13.7 billion years ago,
using our modified gravitational theory without exotic dark
matter. The data fitting ranges over four distance scales: the
solar system, galaxies, clusters of galaxies and the CMB power
spectrum data at the surface of last scattering.

\section*{Acknowledgments}

This work was supported by the Natural Sciences and Engineering
Research Council of Canada. I thank Joel Brownstein, Martin Green
and Justin Khoury for helpful discussions.

\end{document}